\documentclass[prb,twocolumn,showpacs,superscriptaddress,preprintnumbers,floatfix,amsmath,amssymb]{revtex4-1}

\usepackage{amssymb}
\usepackage{graphicx}
\usepackage{dcolumn}
\usepackage{bm}
\usepackage{color}
\begin{document}

\title{Cluster spin-glass ground state in quasi-one-dimensional KCr$_{3}$As$_{3}$}

\author{Jin-Ke Bao}
\email[]{baojk7139@zju.edu.cn}
\affiliation{Department of Physics, Zhejiang University, Hangzhou 310027, China}

\author{Lin Li}
\affiliation{Department of Physics, Hangzhou Normal University, Hangzhou 310036, China}

\author{Zhang-Tu Tang}
\affiliation{Department of Physics, Zhejiang University, Hangzhou 310027, China}

\author{Yi Liu}
\affiliation{Department of Physics, Zhejiang University, Hangzhou 310027, China}

\author{Yu-Ke Li}
\affiliation{Department of Physics, Hangzhou Normal University, Hangzhou 310036, China}

\author{Hua Bai}
\affiliation{Department of Physics, Zhejiang University, Hangzhou 310027, China}

\author{Chun-Mu Feng}
\affiliation{Department of Physics, Zhejiang University, Hangzhou 310027, China}

\author{Zhu-An Xu}
\affiliation{Department of Physics, Zhejiang University, Hangzhou
310027, China} \affiliation{Collaborative Innovation Centre of Advanced Microstructures, Nanjing 210093, China}

\author{Guang-Han Cao} \email[]{ghcao@zju.edu.cn}
\affiliation{Department of Physics, Zhejiang University, Hangzhou
310027, China} \affiliation{Collaborative Innovation Centre of Advanced Microstructures, Nanjing 210093, China}

\date{\today}

\begin{abstract}
We report structural and physical properties of a new quasi-one-dimensional Cr-based compound, KCr$_{3}$As$_{3}$, which is prepared by potassium deintercalation from the superconductive K$_{2}$Cr$_{3}$As$_{3}$. KCr$_{3}$As$_{3}$ adopts the TlFe$_{3}$Te$_{3}$-type structure with space group $P6_{3}$/$m$ (No. 176). The high-temperature magnetic susceptibility obeys the Curie-Weiss law with an effective magnetic moment of 0.68 $\mu_{\mathrm{B}}$/Cr. Below 56 K the susceptibility deviates from the high-temperature Curie-Weiss behavior, coinciding with the rapid increase in resistivity, which suggests formation of spin clusters. The short-range spin correlations are also supported by the specific-heat data. The title material does not exhibit bulk superconductivity; instead, it shows a cluster spin-glass state below $\sim$ 5 K.
\end{abstract}


\maketitle

Superconductivity was recently discovered in a quasi-one-dimensional (quasi-1D) compound K$_2$Cr$_3$As$_3$ (K233) with a superconducting transition temperature $T_{\mathrm{c}}$ of 6.1 K.\cite{bao}  The two sister compounds, Rb$_2$Cr$_3$As$_3$ and Cs$_2$Cr$_3$As$_3$, were subsequently synthesized, which exhibit superconductivity with lowered $T_{\mathrm{c}}$'s of 4.8 K and 2.2 K, respectively.\cite{Rb233,Cs233} The superconducting family is structurally characterized by existence of infinite [(Cr$_3$As$_3$)$^{2-}$]$_{\infty}$ double-walled subnanotubes separated by alkali-metal cations. The monotonic reduction of $T_{\mathrm{c}}$ with the lattice expansion due to incorporation of larger alkali metals suggests an important role of interchain coupling in the occurrence of superconductivity.

Electronic band-structure calculations show that the prototype material K233 hosts a three-dimensional Fermi-surface (3D-FS) pocket in addition to two quasi-1D FS sheets.\cite{caoc} The 3D-FS pocket reflects significant interchain couplings, which may explain the sharp superconducting transitions\cite{bao,canfield} as well as a relatively low anisotropy in the upper critical fields, $H_{\mathrm{c2}}$.\cite{canfield} The normal-state Sommerfeld coefficient is $70-75$ mJ mol$^{-1}$ K$^{-2}$ in K233, over three times larger than that from the band-structure "bare" density of states at Fermi level,\cite{caoc} indicating a significant electron correlation. Unconventional superconductivity was supported by various experimental results such as an extremely large $H_{\mathrm{c2}}$ value exceeding the Pauli paramagnetic limit by a factor over 3,~\cite{bao,canfield} a linear temperature dependence of resistivity above $T_{\mathrm{c}}$ in the polycrystalline samples,~\cite{bao} anomalous spin-lattice relaxation rate behaviors,\cite{imai} and a nearly linear temperature dependence of London penetration depth far below $T_{\mathrm{c}}$.\cite{huiqiu} Contrary to the monotonic increase of $T_{\mathrm{c}}$ with decreasing the lattice dimensions in $A_2$Cr$_3$As$_3$ ($A$ = K, Rb, Cs), pressurizing K233 leads to a monotonic decrease in $T_{\mathrm{c}}$,\cite{canfield,sll} and the $T_{\mathrm{c}}$ suppression rate depends strongly on the pressure-transmitting medium.\cite{sll} Furthermore, ferromagnetic spin fluctuations\cite{caoc} and/or frustrated magnetic fluctuations\cite{hjp1} are concluded based on the first-principles calculations. The theoretical models,\cite{zy,hjp2,djh} built on the basis of molecular orbitals, all suggest unconventional superconductivity with a spin-triplet pairing. The proposed superconducting order parameter has $f$-wave symmetry or $p_{z}$-wave symmetry.\cite{zy,hjp2}

In this Rapid Communication, we report synthesis and characterizations of a related compound KCr$_3$As$_3$ (K133). The new material also contains the (Cr$_3$As$_3$)$_{\infty}$ linear chains, and therefore it can be reasonably regarded as a ``cousin" of the superconductive K233. At first sight, K133 has only one less potassium in the chemical formula. However, the crystal structure differs from that of K233, not only in the lattice parameters but also in point group and space group. Our physical-property measurements indicate that the title compound does not show bulk superconductivity, but exhibits a cluster spin-glass metallic state. The result may shed light on the appearance of superconductivity in K233.

Polycrystalline samples of K133 were synthesized by reacting K233 polycrystals with water-free ethanol [see details in the Supplemental Materials (SM)\cite{SM}]. First, single-phase K233 polycrystals were prepared, as described in the previous report.~\cite{bao} Second, the as-prepared K233 pellets were immersed to the ethanol liquid, holding for five days at room temperature. Third, the resulting pellets were washed by ethanol several times, and then evacuated in vacuum to get rid of the remaining ethanol. It is noted that, unlike K233 which is very air sensitive, the product K133 is stable at ambient conditions.

\begin{figure*}
\includegraphics[width=14cm]{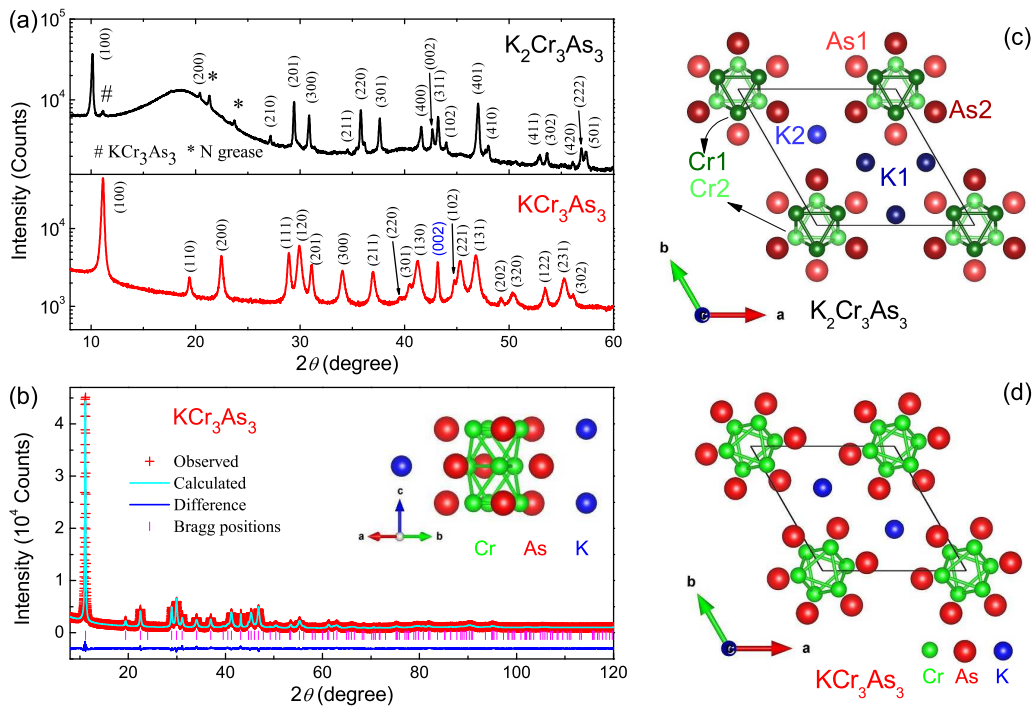}
\caption{\label{XRD} (Color online) (a) X-ray diffraction patterns of K$_{2}$Cr$_{3}$As$_{3}$ (K233) and KCr$_{3}$As$_{3}$ (K133), indexed with hexagonal lattices. Note that a logarithmic scale is applied for the vertical axis to show possible trace impurities. (b) The Rietveld refinement profile for K133. (c) and (d) depict the [001] projection of the crystal structures of K233 and K133, respectively. Different shades of colors in (c) denote different atomic sites in $z$ = 0 and $z$ = 0.5. Nevertheless, the (Cr$_{3}$As$_{3}$)$_{\infty}$ chain, which is more clearly shown in the inset of (b), is similar for both materials.}
\end{figure*}

The powder x-ray diffraction (XRD) data were collected by a PANAlytical x-ray diffractometer (Empyrean) with a Cu$K$$_{\alpha1}$ monochromator (Johansson 1$\times$Ge111 Cu/Co) at room temperature. The Rietveld refinement of the XRD data was performed by the software Rietan-FP.\cite{rietan} Energy-dispersive x-ray spectroscopy was collected by an Octane Plus Detector (AMETEX EDAX), equipped in a field-emitting scanning electron microscope (SEM, Hitachi S-4800). The data were done in several small crystals. The resulted elemental compositions of the K-deintercalated phase corresponds to KCr$_{3}$As$_{3}$ within the measurement errors.~\cite{SM}

Resistivity was measured using a standard four-probe method. Silver paste was used to attach gold wires onto the bar-shape sample. The error of absolute resistivity value is about 20\% due to geometry uncertainty. ac susceptibility and specific-heat data were collected on a Quantum Design physical property measurement system (PPMS-9). For the ac susceptibility measurement, a small oscillated field $H_{\mathrm{ac}}$ = 5 Oe was applied to minimize the influence of field-induced effects. dc magnetic properties were measured on a Quantum Design magnetic property measurement system (MPMS-5). All the measurements employed the same batch of K133 samples.

The K133 ploycrystalline sample is structurally characterized by powder XRD, as shown in Fig.~\ref{XRD}. For comparison, the XRD data of the intermediate product K233 is also presented. As can be seen, after the deintercalation of potassium, the resultant K133 has an obviously different XRD pattern. The small peak marked by \# in the upper plot of Fig.~\ref{XRD}(a) coincides with the strongest peak in the lower plot. This is due to the formation of a small amount of K133 during the XRD experiment (although the sample was coated with a thin layer of N grease). It is noted that the XRD peaks are remarkably broadened for K133. The full width at half maximum is $\sim$ 0.30 degree, almost twice of that for K233. This XRD-peak broadening indicates a reduced crystalline-grain size that could be due to ``fission" of the grains in the process of the topological deintercalation of potassium. Owing to the chain structure with a much stronger intrachain chemical bonding, one would expect a relatively large size along the chain direction. Indeed, the (002) peak is exclusively narrow for K133, suggestive of $c$-axis oriented rod-like crystalline morphology, which can be seen in the SEM observation.\cite{SM}

All the XRD reflections of K133 can be well indexed by a hexagonal lattice with $a$ = 9.0909(8) {\AA} and $c$ = 4.1806(2) {\AA}. The $a$ parameter is reduced by $\sim$ 9\%, and  the $c$ axis is reduced by only $\sim$ 1\%, compared with the unit cell of K233.\cite{bao} This is consistent with the deintercalation of K$^{+}$ in the interchain sites, which makes the chains stack more closely. The $a$ axis is close to that of TlFe$_{3}$Te$_{3}$\cite{TlFe3Te3} whose crystal structure contains similar double-walled subnanotubes. Therefore, we tried a Rietveld refinement using the TlFe$_{3}$Te$_{3}$-type structure with a hexagonal space group of $P$6$_{3}/m$.\cite{TlFe3Te3} The refinement was successful, as shown in Fig.~\ref{XRD}(b), which gives a fairly low reliable factor ($R_{\mathrm{wp}}$ = 3.86\%) and an acceptable goodness of fit ($\chi^2$ = 2.17). The occupancy of K$^+$ is essentially 1.0 after convergence of the refinement. The crystal structure projected along the [110] direction is shown in the inset of Fig.~\ref{XRD}(b). Explicitly, the linear (Cr$_{3}$As$_{3}$)$_{\infty}$ chains remain in K133.

Figures ~\ref{XRD}(c) and (d) display the crystal structures of K233 and K133, projected along the $c$ direction. Compared with K233, the (Cr$_{3}$As$_{3}$)$_{\infty}$ chains of K133 rotate a small angle (along the $c$ axis), such that all the potassium ions are coordinated by nine arsenic anions. The crystallographic data of KCr$_{3}$As$_{3}$ are summarized in Table~\ref{structure}. Unlike K233 where there are two potassium sites, K133 has only one equivalent site for K$^{+}$ ions (regardless of two K$^{+}$ ions in one unit cell), which leads to a different space group of $P6_{3}/m$ with a \emph{centrosymmetric} point group of $C_{6h}$. Owing to the symmetry, all the Cr triangles within the $ab$ plane have exactly the same size in K133. The side length of the Cr triangle, i.e., the Cr$-$Cr bond distance in the basal plane ($d_{\mathrm{Cr-Cr}}$ = 2.691 \AA), is significantly larger than the $d_{\mathrm{Cr1-Cr1}}$ value (2.614 \AA) in K233. This structure change could be relevant to the variations in electronic structure and physical properties, according to the theoretical calculations.\cite{hjp1}

\begin{table}
\caption{\label{structure}
Crystallographic data of KCr$_{3}$As$_{3}$ which crystallizes in a hexagonal lattice with $a$ = 9.0909(8) \AA, $c$ = 4.1806(2) \AA, and space group $P $6$_{3}$/$m$. $g$ denotes the occupancy.}
\begin{ruledtabular}
\begin{tabular}{lccccc}

Atom &Site&$x$&$y$&$z$&$g$\\
\hline
K &$2c$&1/3 &2/3  &0.25&0.998(4)\\
Cr &$6h$ &0.1490(2) &0.1865(2) & 0.25&1 (fixed)\\
As  &$6h$ &0.3407(2) &0.0645(2) & 0.25&1 (fixed)\\
\end{tabular}
\end{ruledtabular}
\end{table}

\begin{figure}
\includegraphics[width=8cm]{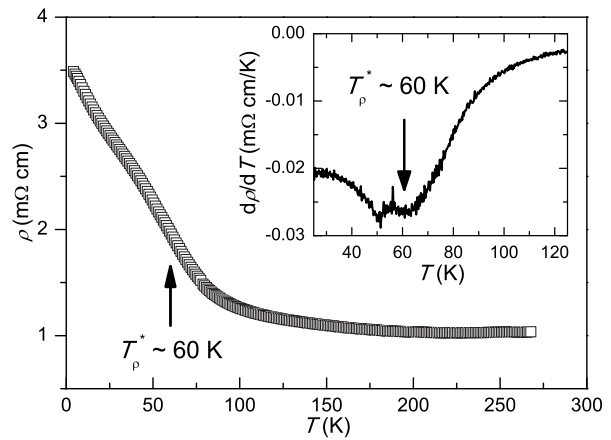}
\caption{\label{rt} Temperature dependence of resistivity for the KCr$_{3}$As$_{3}$ polycrystalline sample. The inset plots the derivative of resistivity.}
\end{figure}

Figure~\ref{rt} shows temperature dependence of resistivity for the K133 polycrystalline sample down to 4 K. The room temperature resistivity is around 1 m$\Omega$ cm, comparable to that ($\sim$ 2 m$\Omega$ cm) of K233.\cite{bao} The resistivity has a very weak temperature dependence above 150 K. Below 150 K the resistivity shows a semiconducting-like behavior; nevertheless, the resistivity at 4 K remains a small value (3.5 m$\Omega$ cm). This ``semiconductivity" seems to be associated with a weak Anderson localization which may be enhanced in the present quasi-1D case. Note that the resistivity increases rapidly at $T_{\rho}^{*}$ $\sim$ 60 K, which is also shown by the valley in ${d\rho}/{dT}$ in the inset of Fig.~\ref{rt}. Below we will see that the characteristic temperature $T_{\rho}^{*}$ coincides with the temperature at which spin clusters form.

\begin{figure*}
\includegraphics[width=14cm]{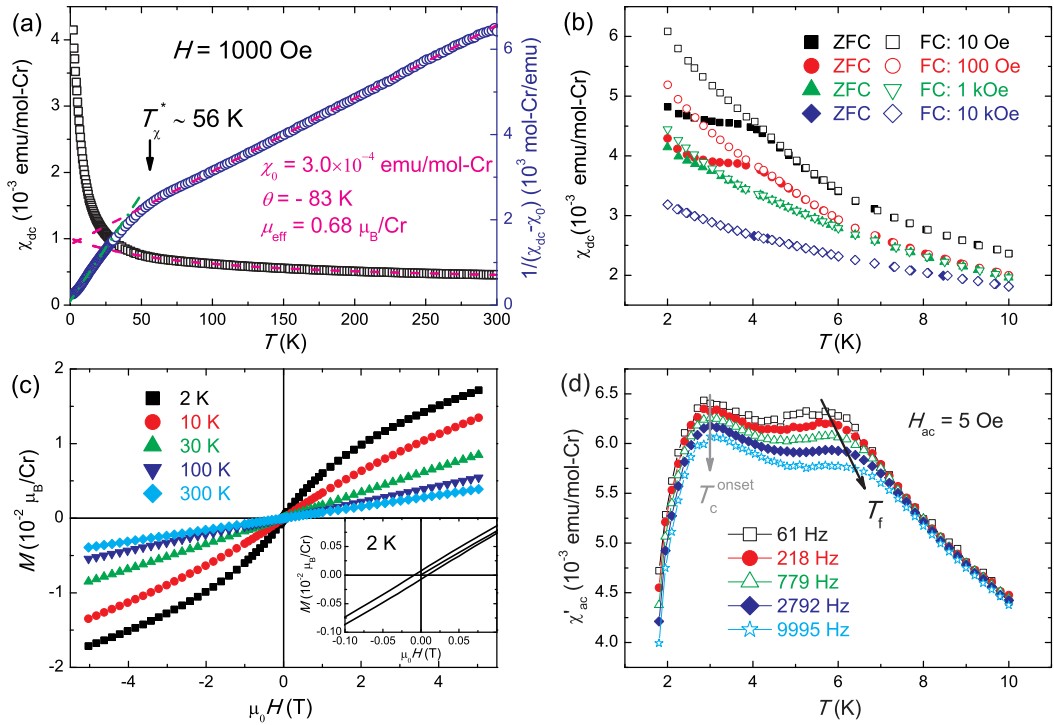}
\caption{\label{magnetic} (Color online) (a) Temperature dependence of magnetic susceptibility for KCr$_{3}$As$_{3}$. The dashed lines represent the Curie-Weiss fits. $T_{\chi}^{*}$ marks the onset temperature where the data deviate from the high-temperature Curie-Weiss behavior. (b) Low-temperature magnetic susceptibility with zero-field-cooled (ZFC) and field-cooled (FC) procedures under different external fields. (c) Isothermal magnetizations at different temperatures. The inset shows the low-field data at 2 K. (d) Real part of ac magnetic susceptibility in different oscillating frequencies at zero dc field. The arrows are guides for the eye.}
\end{figure*}

Figure~\ref{magnetic}(a) shows temperature dependence of dc magnetic susceptibility [$\chi_{\mathrm{dc}}(T)$] of K133. The high-temperature $\chi_{\mathrm{dc}}(T)$ can be well fitted by an extended Curie-Weiss formula, $\chi=\chi_{0}+C/(T-\theta_{\mathrm{CW}})$, where $\chi_{\mathrm{0}}$ is a temperature-independent term, $C$ represents the Curie constant, and $\theta_{\mathrm{CW}}$ denotes the Curie-Weiss temperature. From the fitted value of $C$,  we obtain an effective magnetic moment of 0.68 $\mu_{\mathrm{B}}$/Cr. This suggests that the Cr 3$d$ electrons are partially localized, although most Cr-3$d$ electrons remain itinerant primarily due to the intrachain Cr$-$Cr metallic bonding. Additionally, the fitted Curie-Weiss temperature is $-$83 K, indicating dominantly antiferromagnetic interactions between the Cr local moments. The 1/$(\chi-\chi_{0})$ data have a clear kink below $T_{\chi}^{*}$ = 56 K. The Curie-Weiss fitting of the low-$T$ (7 K$<T<20$ K) data gives a reduced effective magnetic moment of 0.4 $\mu_{\mathrm{B}}$/Cr. This suggests formation of spin clusters. The fitted Curie-Weiss temperature changes to 2 K, revealing a weak \emph{ferromagnetic} interaction between the spin clusters. Note that $T_{\chi}^{*}$ coincides with the $T_{\rho}^{*}$ mentioned above. The rapid resistivity increase may be interpreted by an enhanced magnetic scattering due to the formation of spin clusters.

With further cooling down, a spin-freezing behavior is evidenced. First, $\chi_{\mathrm{dc}}(T)$ shows a divergence at $T_{\mathrm{f}}$ = 4$-$5 K between the zero-field-cooled (ZFC) and field-cooled (FC) data, as plotted in Fig.~\ref{magnetic}(b). The bifurcation tends to close with increasing the external field. Second, the isothermal magnetization curves, shown in Fig.~\ref{magnetic}(c), give additional information to the magnetic behavior. At 100 K and above, the $M(H)$ relation is essentially linear, consistent with the Curie-Weiss paramagnetic behavior. At 10 K, deviations from the linearity are obvious, but no magnetic hysteresis can be detected. At 2 K (below $T_{\mathrm{f}}$), the $M(H)$ curve shows an ``S" shape with a small magnetic hysteresis, suggesting the freezing of spins. Nonetheless, the absolute magnetization is less than 0.02 $\mu_{\mathrm{B}}$/Cr, and it does not saturate at the field up to 5 T, which rules out the possibility of any long-range ferromagnetic ordering. All the phenomena observed above consistently point to a spin-glass state at low temperatures.

The temperature dependence of the real part of ac susceptibility, $\chi_{\mathrm{ac}}'(T)$, further confirms the spin-glass scenario, As shown in Fig.~\ref{magnetic}(d), the $\chi_{\mathrm{ac}}'(T)$ curves display a shoulder above 5 K, which shifts to higher temperatures with increasing the frequency of the applied ac fields. This is one of the hallmarks of spin-glass states.\cite{spin-glass} Note that the drop in $\chi_{\mathrm{ac}}'$ below 3 K, which moves to lower temperatures upon applying a static field (not shown here), is due to trace of the superconducting K233 phase. The superconducting volume fraction is estimated to be less than 0.1\%,  which cannot be detected by the XRD experiment. Owing to the quasi-1D structure, we speculate a filamentary superconductivity that comes from the residual K233 nanorods which are surrounded (and protected also) by the K133 phase.

\begin{figure}
  \centering
  \includegraphics[width=8cm]{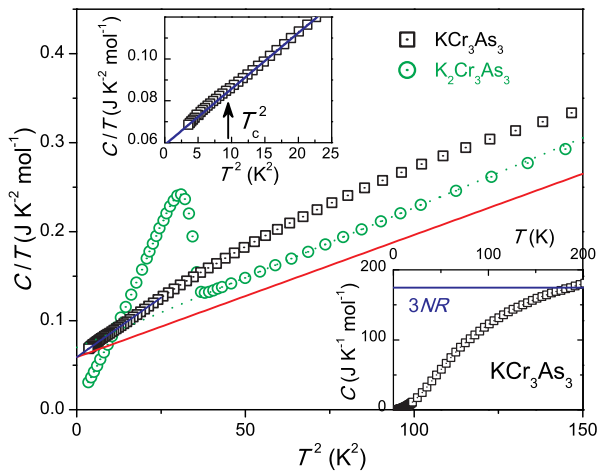}
  \caption{\label{hc} (Color online) Temperature dependence of specific heat for KCr$_{3}$As$_{3}$, plotted in $C/T$ vs $T^2$. The data of superconducting K$_2$Cr$_{3}$As$_{3}$ are presented for comparison. The solid red line approximately represents the expected $C/T$ values without the magnetic contribution. The upper inset is an expanded plot for KCr$_{3}$As$_{3}$, in which an anomaly owing to trace superconducting transition is marked. The lower inset shows the $C(T)$ data in a broad temperature range.}
\end{figure}

Figure~\ref{hc} shows the low-temperature specific-heat data of K133 and K233, plotted in $C/T$ vs $T^2$. In contrast to an abrupt jump at $\sim$ 6 K in K233, no specific-heat jump can be seen for K133. In fact, only a very subtle anomaly appears at 3 K, which can be detected by the deviation of linear fit with the 3 K $<T<$ 5 K data (shown in the upper inset). The magnitude of the small anomaly is less than 1\% of the specific-heat jump in K233, consistent with the above scenario of filamentary superconductivity. In addition, unlike the linear $C/T-T^2$ relation in K233, which gives a Sommerfeld constant of 70 mJ K$^{-2}$ mol-fu$^{-1}$ and a Debye temperature of 215 K, the K133 data obviously deviate from the linearity. This indicates the existence of magnetic contribution ($C_{\mathrm{m}}$), which prevents us from a reliable quantitative analysis.

However, we may have a rough estimation for the $C_{\mathrm{m}}$. First, assuming the same Debye temperature for the two materials (because they have similar chemical components, similar crystal structures and similar average molecular weights), then the lattice contributions to the low-temperature specific heat are related by the formula, $C_{\mathrm{lat}}(\mathrm{K133}) \approx 7/8C_{\mathrm{lat}}(\mathrm{K233})$. Second, the experimental $C/T$ value for KCr$_{3}$As$_{3}$ is 69 mJ K$^{-2}$ mol-fu$^{-1}$ at 1.9 K, which places an upper bound of the Sommerfeld coefficient. The linear fit of $C/T$ vs $T^2$ for the 3 K $<T<$ 5 K data actually gives an intercept of $\gamma_{0}$ = 59 mJ K$^{-2}$ mol-fu$^{-1}$. Third, the high-temperature specific heat obviously exceeds the value (174.6 J K$^{-1}$ mol-fu$^{-1}$) expected from the Dulong-Petit law, as shown in the lower inset of Fig.~\ref{hc}. Taking the same Debye temperature of 215 K, the lattice contribution at 200 K can be calculated to be $C_{\mathrm{lat}}$(200K) = 164.8 J K$^{-1}$ mol-fu$^{-1}$, which is $\sim$15.5 J K$^{-1}$ mol-fu$^{-1}$ lower than the experimental value. The difference should be given by the electronic and magnetic contributions. The magnetic contribution is expected to be relatively small, because the magnetic susceptibility obeys Curie-Weiss law at 200 K. Then, the electronic specific heat, $C_{\mathrm{el}} = \gamma T$, is mainly responsible for the difference. Indeed, if taking the above upper limit of the Sommerfeld coefficient, one can easily get a $C_{\mathrm{el}}$ value of 13.8 J K$^{-1}$ mol-fu$^{-1}$ at 200 K, which is close to the difference. The remaining component, if it exists, could come from the spin correlations which usually extend to high temperatures due to the enhanced fluctuations in quasi-1D systems. Therefore, we draw a straight line by adopting the fitted $\gamma_{0}$ value in Fig.~\ref{hc}, which is believed to approximately represent the sum of $C_{\mathrm{lat}}$ and $C_{\mathrm{el}}$ of K133. With this rough estimation, $C_{\mathrm{m}}(T)$ can be simply extracted by a subtraction, which has a peak at around 12 K (not shown here because it is only qualitatively correct). This phenomenon is common to spin-glass materials in which the spin freezing temperature $T_{\mathrm{f}}$ locates below the peak temperature in $C_{\mathrm{m}}(T)$ and, there is no anomaly at $T_{\mathrm{f}}$.\cite{spin-glass}

Finally, let us make additional remarks on the structural and physical properties of K133. First of all, although the $a$ axis of K133 is greatly reduced, compared with K233, it does not necessarily mean stronger interchain interactions because of different crystal structures (the potassium in the K1 site of K233 may play an important role in the interchain coupling). On the contrary, first-principles calculations suggest a weaker interchain coupling in K133, since only quasi-1D FSs are presented for the calculated ground state.\cite{caoc2} Secondly, K133 shows a Curie-Weiss behavior at high temperatures, indicating existence of local moments. The localized spins do not order as usual, instead, they freeze into a cluster spin glass below 5 K. This result seems to be related to the geometrical frustrations between the Cr local spins and, additionally, to the fluctuations enhanced by the weakened interchain coupling. Thirdly, K133 is basically metallic because of the low resistivity at low temperatures and the appreciable Sommerfeld coefficient. Therefore, the Cr-3$d$ electronic states probably have a dichotomy between local-moment and itinerant scenarios. The result implies that magnetism is important for superconductivity in K233 because of the proximity of the two compounds. Finally, the quench of superconductivity in K133 is apparently related to either the ``hole doping" effect which greatly shifts the Fermi level, and/or the weakening of interchain coupling.\\

\begin{acknowledgments}
We thank C. Cao, J. H. Dai and F. Wang for helpful discussions. This work was supported by the Natural Science Foundation of China (No. 11190023), the National Basic Research Program of China (No. 2011CBA00103), and the Fundamental Research Funds for the Central Universities of China.
\end{acknowledgments}

\end{document}